\documentclass[preprint,12pt]{elsarticle}

\usepackage{amssymb}
\usepackage{graphicx}
\usepackage{multirow}
\usepackage[table,xcdraw]{xcolor}
\usepackage{epstopdf}
\usepackage{amsfonts}
\usepackage{amsmath,amssymb}
\usepackage{color,soul}
\usepackage{tikz}
\usetikzlibrary{automata}
\usepackage{color}
\usepackage{verbatim}
\usepackage{comment}
\usepackage{algorithm}
\usepackage[table,xcdraw]{xcolor}
\usepackage{varioref}
\usepackage{gensymb}
\usepackage{comment}
\usepackage{xfrac}
\usepackage{soul}
\usepackage{color}
\usepackage{xcolor}
\usepackage{gensymb}
\usepackage{comment}
\usepackage{listings}
\usepackage{ifpdf}

\journal{Applied Energy}

\begin{document}

\begin{frontmatter}



\title{ Generation and Evaluation of Space-Time Trajectories of Photovoltaic Power}


\author[NTU]{Faranak~Golestaneh \corref{cor1}}
\author[DTU]{Pierre~Pinson}
\author[NTU]{Hoay~Beng~Gooi}
\address[NTU]{School of Electrical and Electronic Engineering, Nanyang Technological University, Singapore}
\address[DTU]{Department of Electrical Engineering at the Technical University of Denmark, Denmark}
\cortext[cor1]{Corresponding Author: Faranak Golestaneh; Email, faranak001@e.ntu.edu.sg; Phone,  (+65) 6790 4795}


\begin{abstract}
In the probabilistic energy forecasting literature, emphasis is mainly placed on deriving marginal predictive densities for which each random variable is dealt with individually. Such marginals description is sufficient for power systems related operational problems if and only if optimal decisions are to be made for each lead-time and each location independently of each other. However, many of these operational processes are temporally and spatially coupled, while uncertainty in photovoltaic (PV) generation is strongly dependent in time and in space. This issue is addressed here by analysing and capturing spatio-temporal dependencies in PV generation. Multivariate predictive distributions are modelled and space-time trajectories describing the potential evolution of forecast errors through successive lead-times and locations are generated. Discrimination ability of the relevant scoring rules on performance assessment of space-time trajectories of PV generation is also studied. Finally, the advantage of taking into account space-time correlations over probabilistic and point forecasts is investigated. The empirical investigation is based on the solar PV dataset of the Global Energy Forecasting Competition (GEFCom) 2014.

\end{abstract}


\end{frontmatter}


\section{Introduction}
\label{Introduction}
With the increase in the penetration of intermittent generation, a crucial requirement for power systems operation and planning is to enhance forecasting approaches such that they can inform about prediction uncertainties.

 Over the past decade, researchers have intensively investigated solar irradiance and PhotoVoltaic (PV) power point  forecasting for independent sites and forecast horizons.
Time series analysis, AR, ARMA, ARIMA, artificial neural networks, support vector machine are among the mostly used methods in this area~\cite{inman2013solar,marquez2013forecasting,chu2013hybrid,mellit2006adaptive}.
  Point forecast methods work mainly based on minimum least square schemes and can only inform about  conditional expectation of a random variable. Therefore, recently a significant share of practices in the energy forecasting area is concentrated on probabilistic forecasts. These approaches aim at equipping decision makers with appropriate information about stochastic behaviour of the random variables as well as uncertainties attached to the forecasts~\cite{quan2015computational}.
       There are  handful practices on  probabilistic forecasts of  PV generation available in the literature ~\mbox{\cite{lorenz2009irradiance, mathiesen2013geostrophic,chu2015real,alessandrini2015analog}}.
   

If probabilistic forecasts are properly employed, they can serve as a  decision-aiding tool to alleviate challenges attached with stochastic generation. However, despite of the benefits of probabilistic forecasts over point forecasts, they fail to capture development of forecast errors through successive lead-times, interdependent generation in contiguous locations or negatively correlated generation levels in diverse geographic areas~\cite{papaefthymiou2006integration}. The reason is that they treat random variables for each lead-time and each location individually and separately while PV  generations are stochastic processes with spatially spread and time interdependent infeeds. Therefore, in multi-stage decision making problems such as unit-commitment or optimal power flow, it is an integral requirement to estimate aggregated uncertainties in the system and model space-time stochasticity of intermittent resources~\cite{yang2013solar,qin2013incorporating}.

Following complex meteorological mechanisms like cloud passages, PV generations act like a random variable distributed over time and space. Therefore, it is highly plausible that by leveraging spatio-temporal correlations, improved forecast accuracy can be gained.

Just recently, few practices have studied space-time correlations of PV power (or  solar irradiance) and tried to benefit from them in point forecasting. Gueymard and Wilcox~\cite{gueymard2011assessment} have presented a general investigation on long-term variability of solar resources in united states. Yang \textit{et al.} have proposed a statistical approach to obtain temporal and spatial stationarity for solar irradiance time series logged in few numbers of monitoring stations in Singapore. To do so, solar irradiance time series at individual sites are detrended to get temporal stationarity. Spatial stationarity also is obtained by coordinate transformation. Using time forward kriging and with respect  to the persistence method, 25 \%  improvement in RMSE is reported. 

In~\cite{yang2012novel}, spatial cross correlations between all the PV sites under consideration are reported to be more than 0.85. In~\cite{widen2011correlations}, almost the same results are shown where the cross correlations between each pairs of 12 PV zones with distance up to 1500 km in Sweden are found to be more than 0.8. These findings have supported the idea of using data from neighbouring sites as additional explanatory variables for PV generation forecasting at the target location~\cite{yang2015multitime}.

Zagouras \textit{et al.} have looked into space-time correlations of solar irradiance to devise forecast models for seven locations in California with one, two and three hours forecast horizons~\cite{yang2015multitime}. To do so, firstly, the regions which present high correlations between satellite-derived data and ground data are determined. Then, the data for those areas are employed as exogenous variables to predict global solar irradiance at the point of interest. To determine the most optimal time lags for local and exogenous variables, genetic algorithm is used.

In~\cite{bessa2015spatial}, to benefit from spatial-temporal dependencies of PV power, a vector autoregression based method is proposed. Using past observations from the neighbouring locations and for forecast horizons less than 4 hours, up to 10\% improvements in RMSE values are achieved. In~\cite{berdugo2011analog} also to enhance predictability of PV power, measurements from the adjacent PV sites are used as exogenous variables.

As reviewed above, the foundation of the few present studies on the spatio-temporal PV power forecasting is on deploying measurements or meteorological data from the neighbouring locations  in forecasting process of the site of interest (spatial analyses).  Moreover, as the developed methods use lagged data, they are categorized as temporal investigations. Here, though, spatio-temporal correlations of PV power are leveraged in a different way. In contrast with reviewed works in which dependencies are founded on the base of point forecasts, we model the dependencies based on probabilistic forecasts. The goal in this study is to provide more informative forecasts for probabilistic decision making.


 Inspired by the recent multivariate analyses of the wind power, the cornerstone of this practice is to use marginal distributions given by probabilistic forecasts as infeed and couple them using copulas to form a multivariate distribution~\cite{papaefthymiou2009using}.

In~\cite{pinson2009probabilistic}, an approach is proposed to capture and model the time dependent structure of wind power generation for successive hours using marginal densities. With an Gaussian copula assumption, the covariance matrix carries information about temporal dependence. Due to non-stationary characteristics of wind power, an adaptive and recursive method is proposed to track dependency and estimate the covariance matrix for each time. The developed idea is then tested on a multi-MW wind farm with historical data available for a course of two years.


While a range of quantitative assessment frameworks for univariate quantities and independently generated marginal densities exists~\cite{gneiting2007probabilistic}, only few frameworks for the case of multivariate quantity evaluation can be found in the literature. Energy score is the most commonly used scoring rule to evaluate multivariate densities which are described by a finite number of samples. However, the score does not discriminate the misspecified dependency structure between components of a multivariate quantity~\cite{pinson2012evaluating, scheuerer2015variogram}. A variogram-based score is a recently proposed scoring rule and it is claimed to be more sensitive to misspecified mean, variance and correlations~\cite{scheuerer2015variogram}.

An event-based scoring rule is proposed in~\cite{pinson2012evaluating} as an diagnostic approach to assess the correspondence of trajectories generated on a multivariate base and related measurements. Frameworks are provided to evaluate time-dependent trajectories in predicting gradient and long-lasting events.

In this paper, the aim is to investigate and analyse spatio-temporal dependency of PV generations.
Performance of multivariate Normal distribution with both recursive and empirical covariance matrices  is evaluated.
  Quantile regression~\cite{koenker1978regression} is used here to obtain marginal densities independently. These marginal distributions then are employed as infeeds for dependency investigations. Discriminating capability of the relevant scoring rules on performance assessment of space-time trajectories of PV generation has been studied. The scores which originally have been proposed for the case of time-dependency are modified and used for space-time dependency. Three events are proposed for the case of PV generations where PV power measurements for each time of day  are compared with  maximum expected power for the same time. In order to track time dependency of PV generations for successive lead-times while taking into account seasonally variations of sunrise and sunset time, observations and marginal distributions are transferred to a time grid and dependency modelling is carried out on this grid. Eventually, the generating trajectories are transformed back to the original space. The  quality of generated trajectories from multivariate distributions are compared with those drawn from predictive densities, normal distributions centered on point predictions and a generalization of climatology forecasts. Analyses and verification are performed using a dataset including more than two years worth of data with hourly resolution and  three neighbouring PV sites.  
\section{ Experimental Data Description} 
\label{data}

As a basis for  PV generation space-time dependency investigation,  time series of PV generation for three contiguous zones are used. The installation specifications of these zones are given in Table \ref{ZonesData}.

\begin{table*}[t]
\vspace{-0.5em}
	\renewcommand{\arraystretch}{1.3}
\caption{{Description of the three PV zones in this study}}
\label{ZonesData}
\centering
\resizebox{\textwidth}{!}{%
\begin{tabular}{|c|c|c|c|c|c|c|c|}
\hline
             & {\bf Latitude} & {\bf Longitude} & {\bf \begin{tabular}[c]{@{}c@{}}Nominal\\ power\end{tabular}} & {\bf Altitude} & {\bf Panel type}        & {\bf \begin{tabular}[c]{@{}c@{}}Panel Orientation\\ (from the north)\end{tabular}} & {\bf \begin{tabular}[c]{@{}c@{}}Panel\\ Tilt\end{tabular}  } \\ \hline
{\bf Zone 1} & 35\degree$ 16' $$ 30'' $S     & 149\degree$ 06' $$ 49'' $E     & 1560W               & 595m           & Solarfun SF160-24-1M195 & 38\degree Clockwise                                                                      & 36\degree              \\ \hline
{\bf Zone 2} & 35\degree$ 23' $$ 32'' $S     & 149\degree$ 04' $$ 01'' $E     & 4940W               & 602m           & Suntech STP190S-24/Ad+  & 327\degree Clockwise                                                                     & 35\degree              \\ \hline
{\bf Zone 3} & 35\degree$ 32' $S        & 149\degree$ 09' $E        & 4000W               & 951m           & Suntech STP200-18/ud    & 31\degree Clockwise                                                                      & 21\degree              \\ \hline
\end{tabular}
}
\vspace{-0.5em}
\end{table*}
  To predict points and marginal densities of PV generations for three PV zones described above, 12 independent  variables as the output of Numerical Weather Prediction (NWP) provided by European Centre for Medium-Range Weather Forecasts (ECMWF) are used as the explanatory variables. The period for which both NWP and PV measurements are available is from April 2012 until the end of June 2014. Therefore, in total around 800 days worth of data per zone with hourly resolution are used.
  
   The NWP variables employed in this study are  total column ice water, surface pressure, relative humidity at 1000 mbar, total cloud cover, 2 metre temperature, surface solar radiation down, surface thermal radiation  down), net solar radiation at the top of the atmosphere, total precipitation, 10 metre U wind component, 10 metre V wind component. 
  
  Half of the available data is used as the training dataset to generate predictive densities and the second half is used as the evaluation dataset. Dataset has been prepared for the Global Energy Forecasting Competition (GEFCom) 2014 and is available online~\cite{website55}.
\section{Deriving and Evaluating Predictive Marginal Distribution} 
\label{Marginal}

Given PV generation as a multivariate random variable denoted by ${\textbf{P}}_{t} $ with univariate components represented by ${P}_{z,t+k} $ for lead-time $ k $ and location $ z $, and $ F_{z,t+k} $ as univariate strictly increasing cumulative density function (CDF) and $ f_{z,t+k} $ the corresponding density function, $ \alpha $-quantile is the value for which the probability of occurring $ P_{z,t+k} $ below that is equal to $ \alpha_i\in [0,1] $. This function can be mathematically written by
\begin{equation}
\label{eqn_2}
  \Pr(P_{z,t+k}< q_{z,t+k}^{\alpha_i})=\alpha_i \quad \textbf{or} \quad  q_{z,t+k}^{\alpha_i}=F_{z,t+k}^{-1} (\alpha_i)\\
\end{equation}

Using explanatory variables up to time $ t $, a quantile forecast with  nominal proportion $ \alpha_i $ for time $ t+k $ and location $ z $, denoted by $\hat{q}_{z,t+k|t}^{\alpha_i}$ is obtained. As a single quantile contains only a limited information about the likely behaviour of the respective random variable in the future, non-parametric probabilistic forecasts are informed as a set of $ m $ number of quantiles with increasing nominal levels to form a density forecast $\hat{f}_{z,t+k|t}$ as
\begin{equation}
\label{mqunt}
  \hat{f}_{z,t+k|t}=\{\hat{q}_{z,t+k|t}^{\alpha_i }; 0 \leq \alpha_1 \leq \alpha_2 \leq \hdots \leq \alpha_m \leq 1\}    \\
\end{equation}

$\hat{F}_{z,t+k|t}$ is the corresponding marginal predictive CDF and it is derived by looking at each look-ahead time $ k $ and each location $ z $ individuality and independently.

Non-parametric probabilistic forecasts are referred to a group of forecasting methods with no restrictive assumption on the shape and features of the predictive distributions. Quantile regression is the most widely used type of probabilistic forecasting methods introduced in \cite{koenker1978regression}.
 
 Continuous rank probability score (CRPS) is used  to compare the skill of predictive marginals for each component of  $ {\rm{\hat {\textbf{P}}}}_t $, i.e. each lead-time in each contiguous location~\cite{gneiting2007strictly}. CRPS is a negatively-oriented proper score with lower values representing a higher skill of the respective marginal distribution. CRPS compares $ \hat{F}_{z,t+k|t} $ with CDF of corresponding observation $ F_{^{_{z,t + k}}}^0 $. As for each $ t $, $ k $ and $ z $, observation is a single value, the observed CDF is modeled as a single step function when the step from zero to one occurs at the observed value. CRPS for each $ k $ and $ z $ is given by
 \begin{equation}
 \label{CRPS}
 {\rm{CRP}}{{\rm{S}}_{z,k}} = \frac{1}{T}\sum\limits_{t = 1}^T {\left( {\int\limits_0^1 {{{\left( {{{\hat F}_{z,t + k|t}}(p) - F_{^{_{z,t + k}}}^0(p)} \right)}^2}dp} } \right)} 
\end{equation} 
\section{Space-time Dependency Modelling}
\label{Depntsection}

In order to model aggregated uncertainties of a set of concurrent random variables in contiguous or geographically diverse locations, a joint or multivariate distribution is inevitable. Multivariate distributions are usually characterized by a set of trajectories or scenarios drawn from them where the generated trajectories can be treated in the same way as ensemble forecasts are treated  in the realm of methodology. In the following, the derivation of the joint distribution and trajectory generation are outlined. 
\subsection{Multivariate distributions}
\label{multivatesub}

In space-time dependence modelling, we denote PV generation for lead-time $ k $ and  location $ z $ by
\begin{equation}
\label{Ptz}
  P_{t,(z-1)K+k}  \quad  \quad  k=1,2, ..., K, \:
  z=1,2, ..., Z
  \\
\end{equation}

Therefore, the underlying stochastic process is of dimension $ D=K\times Z $ including random variables $ P_{t,1} $, $ P_{t,2} $, ..., $ P_{t,D} $. It should be noted that in this work, upper case letters symbolize random variables while their realizations are expressed by lower case letters. As an example, $ [p_{t,1},  p_{t,2} , ...,  p_{t,D} ]^\top $ denotes realizations of $ P_t $ written as a  vector and the corresponding $ D $-variate CDF, $ \textbf{F}_t $ is described  by 
\begin{equation}
\label{D-CDF}
 \textbf{F}_t(p_{t,1},  p_{t,2} , ...,  p_{t,D})=\Pr(P_{t,1}\leq p_{t,1},P_{t,2}\leq  p_{t,2}, ..., P_{t,D}\leq p_{t,D})\\
\end{equation}
$\hat{ \textbf{F}}_t $ as a predictive multivariate distribution conditional on information up to time t is an estimation of $ \textbf{F}_t $ generated using predictive marginal distribution for each random variable ${F}_{t,d} \: (d=1,..., D) $.

Based on Sklar's theorem~\cite{sklar1959fonctions}, to model the interdependence structure of a set of random variables, respective marginal distributions can be linked together using a copula function as
\begin{equation}
\label{coplas}
 \hat{\textbf{F}}_t(p_{t,1}, p_{t,2} , ..., p_{t,D})=C(\hat{F}(p_{t,1}),  \hat{F}(p_{t,2}) , ...,  \hat{F}(p_{t,D}))\\
\end{equation}

A copula is defined as a multivariate distribution function  for which the marginal  distribution of each random variable is uniform on $ [0,1] $. Therefore, in order to use copulas, random variables $ P_{t,d} \:(d=1,...,D) $ should be transformed to a uniform domain, also called rank domain. Then, dependency modelling is carried out in this common domain. 

 By definition of a CDF for a continuous variable, $ \hat{F}_{d}(P_{d}) $ is distributed uniformly as $ \hat{F}_{d}(P_{d})\sim  U[0,1]$. Therefore, to transform random variables to a uniform domain, a simple CDF transformation can be used. The transformed variables preserve the dependency structure which is presented in the original domain~\cite{nelsen2013introduction}. It is noteworthy that continuous CDFs can be obtained by fitting a smooth curve  through  the set of quantiles available.  As CDF is a strictly increasing function, the variables in the uniform domain can be transferred back to their original domain later using the inverse CDF transformation. This transformation can be mathematically written as
 \begin{equation}
 \label{transofrm}
  P_{d}= \hat{F}_{d}^{-1 }(U)  \Leftrightarrow U=\hat{F}_{d}(P_{d})\\
 \end{equation}
 
 Accordingly, a copula function can be represented by
  \begin{multline}
  \label{Copla}
   C(U_1,U_2, ..., U_D)= \Pr(\hat{F}_1(P_1)\leq U_1, \hat{F}_2(P_2)\leq U_2,...,\hat{F}_D(P_D)\leq U_D)\\
   =\textbf{F}(\hat{F}_1^{-1}(U_1), \hat{F}_2^{-1}(U_2),...,\hat{F}_D^{-1}(U_D))
  \end{multline}
  
  The formulation for Gaussian, T and Gumbel copulas can be found in~\cite{nelsen2013introduction,demarta2005t}. Unfortunately, there is almost no guidance available on which copula family can describe correlated variations in PV power. For the wind case though, some practices have investigated the performance of several families of copulas like Archimedean and Elliptical copulas~\cite{louie2014evaluation}. To model wind power for systems with only few variates, Gumbel copula is recommended as the right one to be used. However, as the dimension of the system increases, more probably Gaussian copula can outperform Gumbel one~\cite{diaz2014simulation,diaz2014note} Here, the focus is given to  Gaussian copula because based on our empirical tests discussed in section \ref{Results}, it is able to provide adequate superiority over the benchmarks introduced. However, studying the right copula is essential piece research that should be followed by future works in this area.

In the following subsections, with the focus on Normal Copula, the multivariate distribution derivation and space-time trajectory generation are explained. 
\subsection{Multivariate Normal distribution with adaptive correlation matrix}

Given $ \hat{F}_{t,d} $, random variable $ Y_d $ with $ y_{t,d} $ as its realization at time $ t $ can be defined by 
 \begin{equation}
 \label{Y}
    y_{t,d}= \hat{F}_{t,d}(p_{t,d}) \quad \forall t \\
 \end{equation}  
\lowercase{where}  $ Y_d $ is uniformly distributed on interval $ [0,1] $. Given $ Y_d\sim U[0,1] $, a normally distributed $ X_d $ can be attained using the probit function $ \Phi^{-1} $ as
 \begin{equation}
  x_{t,d}=\Phi^{-1}(y_{t,d}) 
   \end{equation}
\lowercase{where} the probit function is 
   \begin{equation}
   x_{t,d}=\sqrt{2}\:\text{erf}^{-1}(2y_{t,d}-1) 
   \end{equation}
  If the forecasted CDFs are calibrated, the random variable $ \{ x_{t,d}\}_{t=1}^T $ for  each $ d $ follows the standard Gaussian distribution, i.e. $ X_d\sim\mathcal{N}(0,1)$. 
  
   Denoting $ \textbf{X} =(X_{1},X_{2},...,X_{D})^\top $, $ \textbf{X}$ follows a multivariate Gaussian distribution,  $ \textbf{X}\sim\mathcal{MVN}(\mu_0,\Sigma) $, where $ \mu_0 $ is a vector of zeros and size $ D $ and $ \Sigma_t $ is a variance-covariance matrix of size $ D\times D $. 
   
   If historical data for a long period of time is available, the empirical correlation matrix can be calculated.   Alternatively, an adaptive approach can be used to estimate $ \Sigma$ for each time $ t $ as
  


     \begin{equation}\label{5}
                  \Sigma_t = \lambda\Sigma_{t-1} + (1-\lambda)\:\: \textbf{X}_t \textbf{X}_{t}^{{\top }}  \quad \forall t>1
     \end{equation}
 \lowercase{where} $ \lambda $ is  a forgetting factor ($ \lambda\in[0,1) $). $ \Sigma_t $ is initialized (for $ t=1 $) by setting the diagonal elements equal to unity and all the other elements equal to zero.
 
 With the assumption of perfect reliability of predictive quantiles, the diagonal elements of $ \Sigma_t $ should remain equal to 1 through updating steps. However, due to deviations from perfectly calibrated quantile forecasts, they may deviate from 1. If it is assumed that such deviations are caused by variance scaling, the following transformation can be used to make the covariance matrix a proper representative for a standard Gaussian variable. 
 \begin{equation} \label{diagnal}
                      \Sigma_t =\Sigma_t \varnothing (\sigma_t\sigma_t^{\top }) 
 \end{equation}
\lowercase{where} $ \sigma_t $ is a standard deviation vector of size  $ D $ containing the square root of the diagonal elements of $ \Sigma_t $ and $ \varnothing $ is an operator representing the element-by-element division.
        
 The correlation matrix should be positive semi-definite. However, due to noisy estimation of the correlation matrix in either of the approaches (recursive or fully empirical), $ \Sigma$  often appears to be nonpositive semi-definite. Therefore, it is necessary to adjust the violation by approximating $ \Sigma$ to the most closely positive semi-definite matrix. 
\subsection{Trajectory generation}
\label{Trajectory Generation}
To generate several scenarios of the possible values that the random variable $ P_{t,d} $ may take, the most up-to-date covariance matrix which can be deployed is $ \Sigma_{t-1} $. It is to be noted that in this practice, $ t $ is the day index. Given the forecasted $\hat{f}_{t,d}$ and $\hat{F}_{t,d}$ for all involved random variables, and the multivariate distribution of $ \textbf{X}$, $ \textbf{X}\sim\mathcal{MVN}(\mu_0,\Sigma_{t-1}) $, $ S $ scenarios are generated as follows

\begin{itemize}
  \item Draw $ S $ realizations of $ \mathcal{MVN}(\mu_0,\Sigma_{t-1})$. Each of them is a vector of size $ D $, denoting  $ \textbf{x}_{s,(t-1)}$
  \item The elements of the vector of uniform variable $ \textbf{Y}_s=\{{y_{s,d}}\}_{d=1}^D  $ for scenario $ s $ is achieved by employing the inverse of the probit function as 
   \begin{equation}
                       y_{s,d} =\Phi(x_{s,d}) \quad \forall s,d
   \end{equation}
  
  \item The probable scenarios of $p_{s,t,d}$ can be obtained by
  
    \begin{equation}
                       \hat{p}_{s,t,d}  =\hat{F}_{t,d}^{-1}(y_{s,d}) \quad \forall s,d
    \end{equation}
  
\end{itemize}
  \lowercase{where} $  [\hat{p}_{s,t,1},\hat{p}_{s,t,2}, ..., \hat{p}_{s,t,D}] $ represents the $ s^{th} $ space-time trajectory of PV generation for time $ t $.
\section{Multivariate Scoring Rules} 
\label{Scoring Rules}
 The benefits offered by  probabilistic forecasting  to decision making process are directly dependent on the quality of forecasts.  While a range of univariate scoring rules can be found~\cite{gneiting2007probabilistic, pinson2007non}, only a few quantitative assessment criteria for multivariate setting have been proposed.  Some of the most relevant scoring rules for evaluation of the multivariate predictive trajectories for the case of spatially-temporally correlated PV generations are used here.
 \subsection{Energy score}

The most commonly used scoring rule when distributions are represented by a finite number of trajectories is known as energy score (ES). ES as a multivariate generalization of the CRPS has been formulated and introduced in~\cite{gneiting2007strictly}. This score is proper and negatively oriented, i.e. a lower score represents a better forecast. Energy score is calculated as
\begin{equation}
\label{Est}
{\rm{E}}{{\rm{S}}_{\rm{t}}} = \frac{1}{S}{\sum\limits_{s = 1}^S {\left\| {{{\rm{\textbf{p}}}_t} - {\rm{\hat {\textbf{p}}}}_t^{(s)}} \right\|}_2} - \frac{1}{{2{S^2}}}{\sum\limits_{s' = 1}^S {\sum\limits_{s = 1}^S {\left\| {{\rm{\hat {\textbf{p}}}}_t^{(s')} - {\rm{\hat {\textbf{p}}}}_t^{(s)}} \right\|}_2 }}
\vspace{-0.4em}
\end{equation} 
\lowercase{with}  $ {\rm{\hat {\textbf{p}}}}_t^{(.)} $ as trajectories distributed according to the predictive multivariate CDF $ \hat{\textbf{F}}_{t} $, and $ \left\| . \right\| $ is the $ K\times Z $ dimensional Euclidean norm. ES is averaged over the $ T $ number of forecast time series. 

ES has a good discriminating ability to evaluate forecasts relying on marginals with correct variances but biased means. However, it is insensitive to misspecification of dependence structures~\cite{pinson2012evaluating} and weak at discriminating multivariate forecasts when correlations between their components make the only distinction between them~\cite{scheuerer2015variogram}.
 \subsection{Variogram-based score}
 Due to the limitations of ES on detecting misspecified dependence structures between elements of a multivariate quantity, Variogram-based Score (VS) as a proper score is introduced in~\cite{scheuerer2015variogram}. VS is discriminating on misspecified means, variance and correlations. For a $ d $-variate observation and $ \hat{\textbf{F}}_{t} $, VS of order $ \gamma $ can be written as   
\begin{equation}
\label{VSt}
V{S_t} = {\sum\limits_{i,j = 1}^d {{w_{ij}}\left( {|{p_{t,i}} - {p_{t,j}}{|^\gamma} - {{\rm{E}}_F}|{\rm{\hat {\textbf{P}}}}_{t,i} - {\rm{\hat {\textbf{P}}}}_ {t,j}{|^\gamma}} \right)} ^2}
\end{equation} 
\lowercase{where} $ {\rm{\hat {\textbf{P}}}}_{t,i} $ and $ {\rm{\hat {\textbf{P}}}}_{t,j} $ are the $ i^{th} $ and $ j^{th} $ element of random variable  $ {\rm{\hat {\textbf{P}}}}_t $ distributed according to $ \hat{\textbf{F}}_{t} $ for which the $ \gamma^{th} $ absolute moment exists. With a set of trajectories as a representative of $ \hat{\textbf{F}}_{t} $, forecast variogram can be approximated by
\begin{equation}
\label{VSfor}
{{\rm{E}}_F}|{\rm{\hat {\textbf{P}}}}_{t,i} - {\rm{\hat {\textbf{P}}}}_ {t,j}{|^\gamma } \approx \frac{1}{S}\sum\limits_{s = 1}^S {{\rm{|\hat p}}_{t,i}^{(s)} - {\rm{\hat p}}_{t,j}^{(s)}{|^\gamma }} , \quad \forall i,j = 1,2,...,d.
\end{equation}
$ V{S_t} $ then is averaged over the time series available in the evaluation dataset as
\begin{equation}
\label{VSavrg}
VS = \frac{1}{T}\sum\limits_{t = 1}^T {V{S_t}}
\end{equation} 

It is discussed in~\cite{scheuerer2015variogram} that down-weighting pairs of components of  $ {\rm{\hat {\textbf{P}}}}_t $ which are expected to have relatively low correlations would be helpful to improve signal to noise ratio and alleviate sampling error. As it is expected that pairs of the elements with higher distance or longer intervals to have lower relative correlations, pairwise $ {w_{ij}} $ can be defined proportional to inverse distance between   $i^{th} $ and $ j^{th} $ components of $ {\rm{\hat {\textbf{P}}}}_t $. 
\subsection{Event-based score}
\label{eventscore}
The scores mentioned above are helpful when the goal is to discriminate and compare the quality of different sets of trajectories. However, they do not identify and inform about the ability of a set of forecasts in mimicking the stochastic process and predicting the likelihood of the events occurring in the concerned process. An event-based score is proposed in~\cite{pinson2012evaluating} to fill this gap. This score measures the capability of a set of predicted  trajectories for predicting long-lasting and gradient events. In~\cite{pinson2012evaluating}, long lasting events are defined as the successive hours with wind generations continuously more than a predefined threshold. A gradient event also is described as the maximum  absolute ramp which has realized or predicted to occur over a window of size $ h $ and centred on the $ k^{th} $ lead-time. 

There is a fundamental difference between wind  and PV variations. Changes in PV generations can be divided into fairly predictable variability and uncertainty. Variability of PV power is the consequence of earth rotation around its own axis and can be described by clear sky solar irradiance. However, the intermittency attached to PV power is caused by various factors such as cloud passages. Even   with clear sky and no sources of intermittency, it is not expected that PV generations stay in the same level for successive hours. This variability, for the solar irradiance is highly predictable and can be considered as deterministic variations, although when it comes to PV generation the accuracy of estimation would be lower. Therefore, different with the case of wind power, for PV generation each time of day should be treated different from the others and dependent on the maximum expected power for that time. Maximum expected power can be assumed to be the expected non-overcast PV generation. Depending on the available information for the location of interest,  various models can be found to estimate clear sky solar irradiance. There is no perfect model for solar irradiance to PV power conversion though. Parameters such as shadow effect of building and trees, dust on the panels, maximum power point tracking system, power electronic circuits connected to the PV system  cannot be modelled in a straightforward way. Therefore, inspired by~\cite{bacher2009online}, quantile regression as a statistical smoothing method is deployed here. The quantile with the coverage level of 99\% is regarded as the maximum expected PV generations. Then, for event-based analyses, the intermittency with respect to the assumed maximum expected power is measured. 

For the PV generation, three events are proposed in this study.
\subsubsection{Event 1: Highly intermittent hours}
This event describes the proportion of  daytime when the difference between measurements and maximum expected power exceeds a predefined threshold. Those periods are regarded  as highly intermittent and less predictable hours.  Comparing to less intermittent periods, during those hours more reserve or flexible generation are required  to compensate fluctuations. This event can be mathematically written by 

\begin{equation}
\label{intermittent hours}
g_{t}\left( {{\textbf{p}_t};{\textbf{pc}_t},k,h,\xi } \right) = \coprod\limits_{i = k - {h/2}}^{i = k + {h/2}} {\textbf{1}\left\{ {{|pc_{z,t + i}-p_{z,t + i}|} \ge \xi } \right\}} \quad  z = 1,...,Z. 
\end{equation} 
\lowercase{where} $ pc_{z,t + i} $ denotes maximum expected power for time $ t $, zone $ z $ and step size $ i $. $ \xi $ is the threshold value and $ h $ is the window size. $ k $ is the center of the window and so can be considered as a form of look-ahead time. All these three parameters are application dependent and decided by the decision-maker. $ {\textbf{1}\left\{ {} \right\}} $ is an indicator variable with value of one if the condition realizes and zero value otherwise.
\subsubsection{Event 2: Long-lasting  intermittent hours}
The difference between this event and event 1 is that, event 2 monitors the periods with intermittency continuously more than a predefined thresholds. The formulation is given by

\begin{equation}
\label{eventlong}
g_{t}\left( {{\textbf{p}_t};{\textbf{pc}_t},k,h,\xi } \right) = \prod\limits_{i = k - {h/2}}^{i = k + {h/2}} {\textbf{1}\left\{ {{|pc_{z,t + i}-p_{z,t + i}|} \ge \xi } \right\}} \quad  z = 1,...,Z. 
\end{equation}
\subsubsection{Event 3: Gradient predictability}
If the difference between maximum and minimum deviations of measurements from the maximum expected power over a window of size $ h $ and centered on the $ k^{th} $ exceeds a predefined threshold, this status can be regarded as a gradient event. A gradient event can be given by

\begin{align}
\label{gradint}
g_t\left( {{\textbf{p}_t};k,h,\xi } \right) = \textbf{1}\left\{ {\left( {\mathop {\max }\limits_{i \in \left\{ {k - {h/2},...,k + {h/2}} \right\}} {|\Delta_{z,t + i}|}
 - \mathop {\min }\limits_{i \in \left\{ {k - {h/2},...,k + {h/2}} \right\}} {|\Delta_{z,t + i}|}} \right) \ge \xi } \right\}
\end{align}
\lowercase{with}
\begin{align}
\label{gradin}
\Delta_{z,t + i}=pc_{z,t + i}-p_{z,t + i}
\end{align}
Other related events can be formulated depending on  requirements of the decision maker. For example, in gradient events, one might be more concerned about the minutes during the noon time when PV generation drops below a predefined threshold. 

Eq. \eqref{intermittent hours} to \eqref{gradint} represent events in an observed trajectory. To calculate the proportion of the trajectories predicting the concerned event \eqref{gradin2} can be used. 
\begin{equation}
\label{gradin2}
\hat g_t\left( {\hat {\textbf{p}}_t, {\textbf{pc}_t};\theta } \right) = \frac{1}{S}\sum\limits_{s = 1}^S {g_t\left( {\hat {\textbf{p}}_t^{(s)},{\textbf{pc}_t};\theta } \right)} 
\end{equation}
\lowercase{where}  $ h $, $ k $, and $ \xi $ are substituted by  parameter set $ \theta $. 

Brier score (BS) as a proper scoring rule then can be employed to measure the quadratic difference between probability forecast of a set of trajectories and observed one for a defined event by
\begin{equation}
\label{Bs}
{\rm{BS}} = \frac{1}{T}{\sum\limits_{t = 1}^T {\left( {\hat g\left( {\hat {\textbf{p}_t}^{(s)};\theta } \right) - g_t\left( {{\textbf{p}_t};\theta } \right)} \right)} ^2}
\end{equation}
\subsection{Probability Integral Transformation \textup{(}PIT\textup{)}}
To evaluate the probabilistic calibration of generated trajectories with respect to marginal predictive distributions, PIT histograms  can be employed. In PIT histograms, the percentage of generated scenarios lying between two successive quantiles of marginal distributions are compared with the  difference between nominal probability of those quantiles  (known as ideal case).

\section{Results}
\label{Results}
In order to get an insight about the aggregated uncertainties in the system, it is crucial to model the related dependencies.  The database described in section \ref{data} is used to investigate dependency structures existing in PV generations.  Time-dependency is studied here by looking at zone 1 only, hence neglecting  correlations in space. Space-time correlations are also investigated by treating all three zones and forecast horizons simultaneously. The forecast horizons in this practice are 6 to 19 hours ahead.  It is to be noted that NWP for the dataset used in this paper are issued at midnight. For instance when predicting PV generations for 6 am and 12 pm, the forecast horizons are 6 and 12 hours respectively. 

\begin{figure*}[h]
\vspace{-1.2em}
		\begin{center}
		\centering
		\noindent
		\includegraphics[width=10cm]{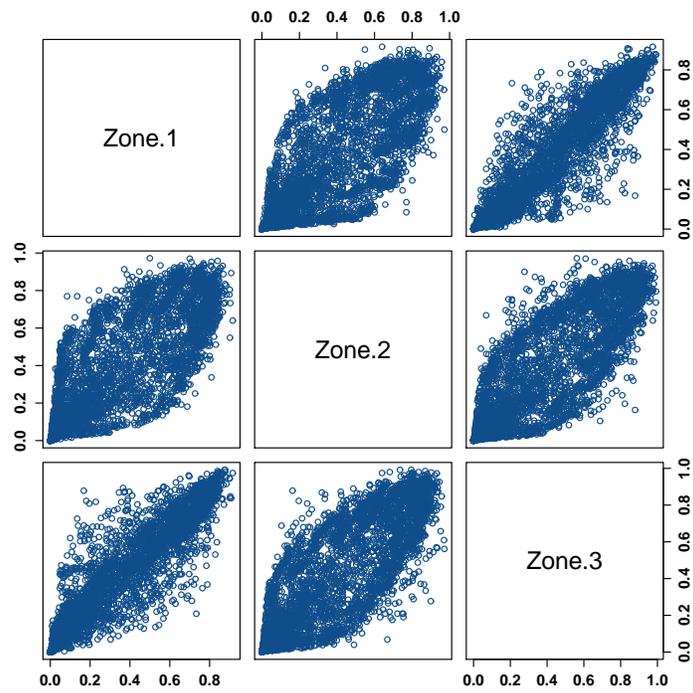}
	\end{center}
	\vspace{-1.2em}
	\caption{All pairwise scatter plots of PV measurements in three zones available in the evaluation dataset }
	\label{scatter}
	 \vspace{-0.4em}
\end{figure*}

To get an idea of how PV generations for the three zones used in this practice are correlated in space, a pairwise scatter plot of historical data is provided in Fig. \ref{scatter}. From Fig. \ref{scatter}, it can be observed that pairwise zones are highly and positively correlated. PV generations for diverse geographical regions can also be negatively correlated. This means that for instance if the western part of a region  is rainy, there is a certain probability that the eastern part is experiencing a sunny time.

\subsection{ Benchmark methods}
\subsubsection{Naive method}
A natural benchmark here would be an analogy-like approach. In this method, trajectories are randomly selected from the past observations. For instance, if the number of required trajectories is $ S $, the observed times series for $ S $ randomly selected days from the past act like the set of predicted time-dependent trajectories for the current day. It is like assuming that one possibility is that what happens today is exactly the same as one of the previous days. If the number of trajectories was tending towards infinity, one would end up with a generalization of climatology forecast in the multivariate context.
\subsubsection{Independent method}
In this benchmark, for each $ t $, $ k $ and $ z $, trajectories are generated by uniformly sampling from the corresponding predictive marginal distribution. Therefore, this benchmark takes advantage of probabilistic forecasts though neglects space-time correlations.
\subsubsection{Independent Gaussian}
In probabilistic decision-making approaches such as Monte Carlo based methods, usually it is assumed that forecast errors follow a normal distribution centered on point forecasts. This benchmark is considered here to investigate the benefits of probabilistic forecasts over the common practice of normality assumption for  uncertainty attached with intermittent generations. In this benchmark, trajectories for each $ t $, $ k $ and $ z $ are generated by sampling from a normal distribution centred on corresponding point predictions.

In order to find standard deviation  values which can reasonably describe uncertainties around point predictions, two options are tried.
In the first option, the standard deviation for each $ t $, $ k $ and $ z $ is calculated using historical data. For the second option, it is assumed that standard deviation for each $ t $, $ k $ and $ z $ is $ \beta \%$ of the corresponding point prediction. As the calculated scores for the second option found to be much better than the first one, here only results for the second option are given. Two common assumptions for the value of  $ \beta $ are 5 and 10. The same values are chosen here  for independent Gaussian-5 \% and independent Gaussian-10 \% respectively. It is to be noted that point forecasts are generated by deploying the method proposed in~\cite{Batch2015}.
\subsubsection{Gaussian copula with full empirical covariance}

This benchmark is considered to make a comparison between multivariate normal distributions with the recursive covariance matrix and empirical one. Hence, in benchmark 5, covariance matrix is calculated using historical data available up to time $ t $.  

\subsection{Time grid}
High dependency of PV generations on seasonally sunshine variations enforces a major difference between time dependent wind and PV generations. Given data with hourly resolution, sunrise and sunset vary seasonally (or monthly).  In order to keep the size of the covariance matrix fixed and at the same time take into account sunrise and sunset changes, one approach is to project the information originally available (of varying dimension, since the beginning and end of day may change) onto a grid called time grid whose dimension stays fixed. For each day, the first element of the corresponding grid is the first PV measurement after sunrise and the last one is the last measurement before the sunset. The rest of the elements are found by interpolation among values for daytime hours. Then, for all the covariance modelling, only information on this grid is considered. Eventually, when generating trajectories, they have to be transformed back to the original space and skill verifications are performed in the original space. The size of the grid is chosen to be 15 while there is no restriction on selecting a higher grid size.
\subsection{Marginal distributions}
\begin{figure}[h]
         \vspace{-1.2em}
         		\begin{center}
         		\centering
         		\noindent
         		\noindent\makebox[\textwidth]{
         		\includegraphics[width=15cm]{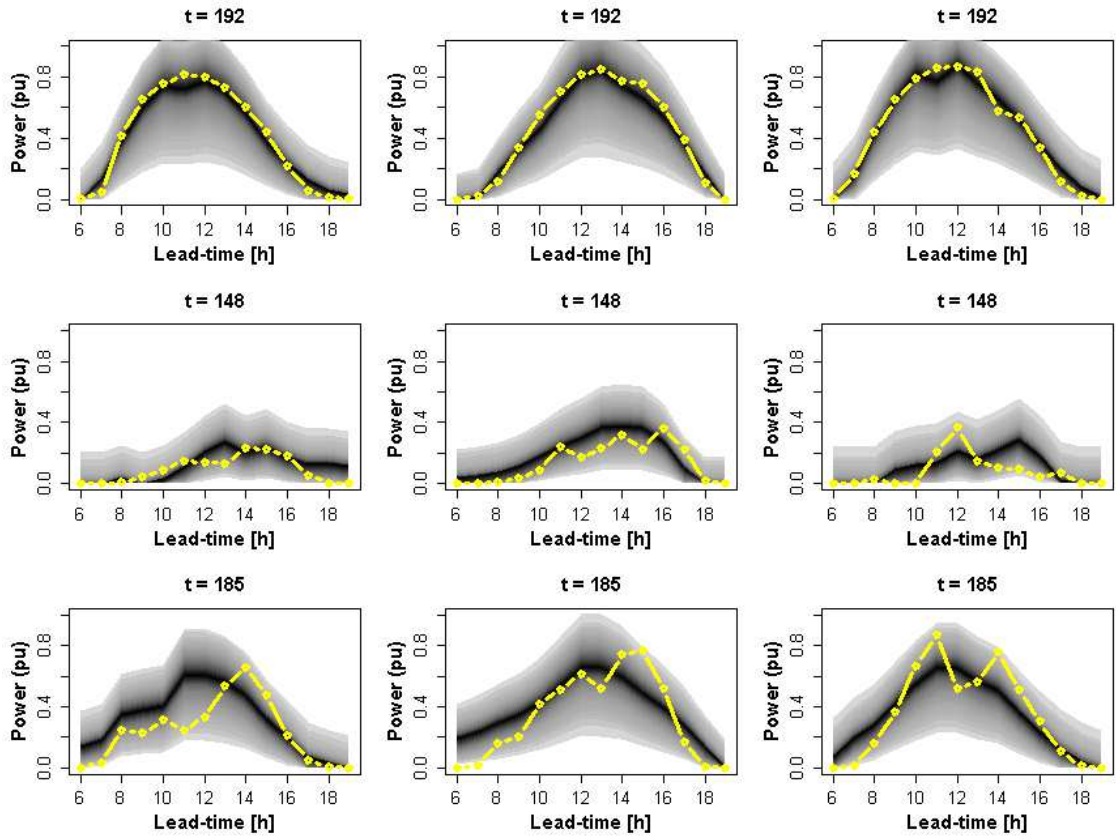}
         		}
         	\end{center}
         	\vspace{-1.2em}
         	\caption{PV observations (yellow colour curves) along with 49 prediction intervals with coverage ranging from 0.02 to 0.98 by 0.02 increments (from the darkest to the lightest), for 3 randomly selected days from the evaluation set, column 1 represents data for zone 1, column 2 data for zone 2 and column 3 data for zone 3}
         	\label{PIAllzones}
         	 \vspace{-0.4em}
   \end{figure}

Fig. \ref{PIAllzones} illustrates probabilistic forecasts in the form of non-parametric predictive distributions obtained using quantile regression~\cite{koenker1978regression} for the three randomly selected days from the evaluation dataset. Univariate predictive distributions are described by 99 quantile forecasts with nominal coverage ranging from 0.01 to 0.99 with 0.01 increments. Then, as by products of quantile forecasts, 49 prediction intervals with nominal proportion from 0.02 to 0.98 with 0.02 increments are generated. Each prediction interval is central on median and is generated by using two quantile forecasts with symmetrical nominal probabilities with respect to the median. 12 independent NWP variables described in section \ref{data} are used as the predictors and PV power is treated as the output of the quantile regression.

\begin{figure}[!t]
     	\vspace{-0.4em}
     	\centering
     	\includegraphics[width=8cm, height=7cm]{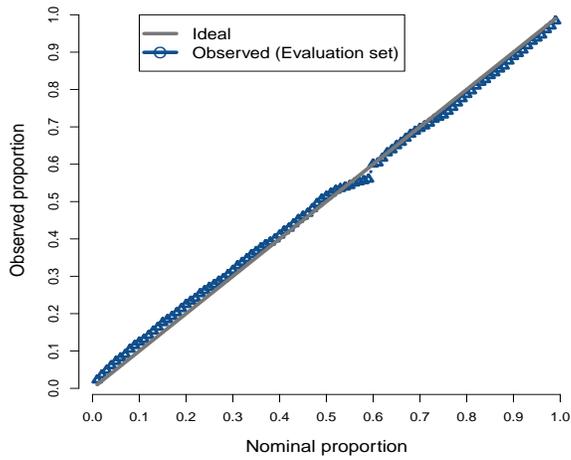}
     	\caption{Relibility evaluation of 99 quantiles with 0.01 increasing nominal proportion generated by quantile regression for zone 1 and all forecast horizons.}
     	\label{QQmarginal}
     	\vspace{-0.6em}
     \end{figure}

In order to benefit from quantile forecasts, they should be calibrated. Calibration of the quantile forecasts can be verified via reliability diagram in which the proportion of the measurements with values less than each quantile of the marginal distribution is compared with the nominal probability of the corresponding quantile (ideal case). Fig. \ref{QQmarginal} illustrates reliability diagram for zone 1 when all forecast horizons are covered. As can be seen, the deviations from the ideal reliability are quite low and that lets us to use the generated marginal distributions for further investigations.

Skill verification of predictive marginal distributions based on CRPS is visualized in Fig. \ref{CRPS} for all three zones available. From the figure, it can be concluded that for zones 1 and 3, marginal densities are less skilful around 9 am in the morning while zone 2 has its peak in the afternoon. However, for all zones it can be seen that CRPS mimics the typical daily sunshine curve with lowest values close to sunrise and sunset.
\subsection{Multivariate analysis}
\begin{figure}[h]
         \vspace{-1.2em}
         	\begin{center}
         		\centering
         		\noindent
         		\noindent\makebox[\textwidth]{
         		\includegraphics[width=15cm]{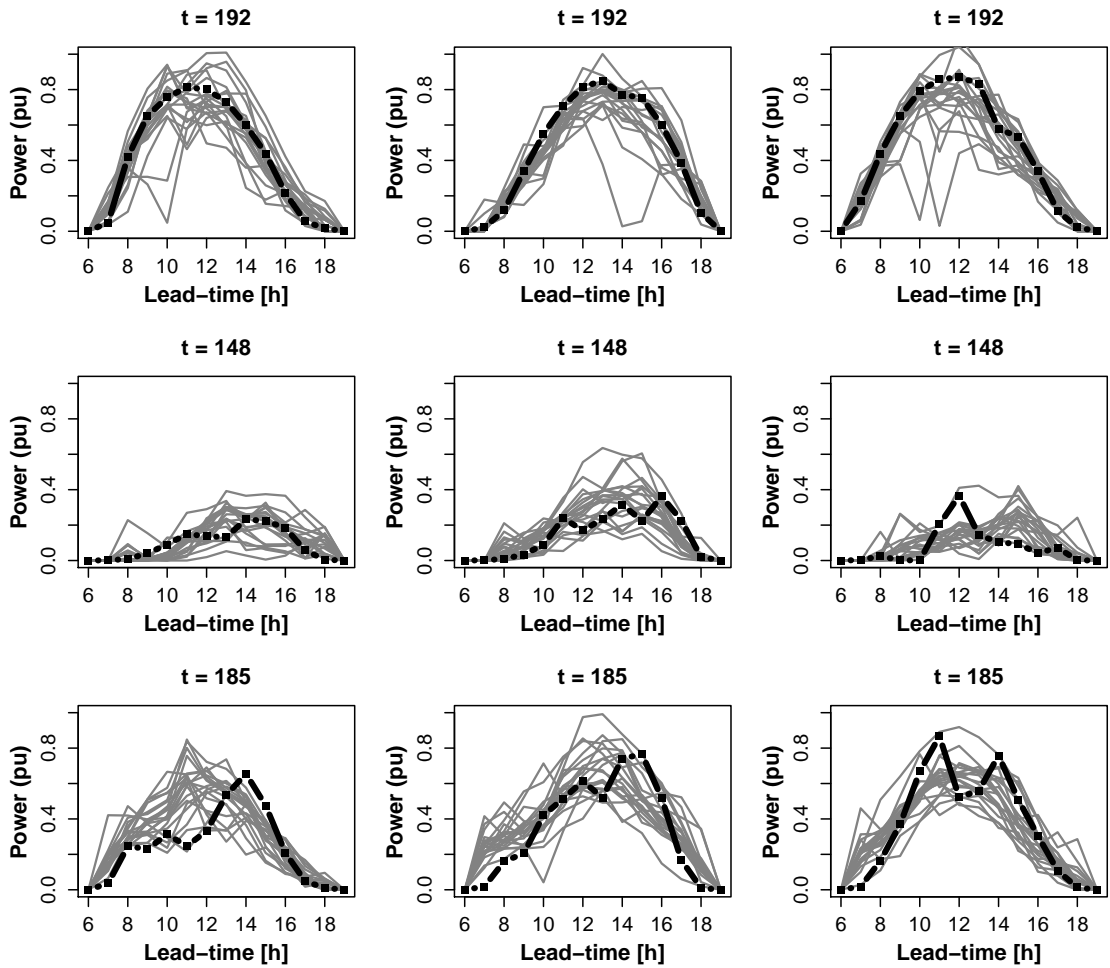}
         		}
         	\end{center}
         	\vspace{-1.2em}
         	\caption{ PV observations (dark black colour curves) along with 20 generated space-time trajectories (grey colour curves) for 3 randomly selected days from the evaluation set, column 1 represents data for zone 1, column 2 data for zone 2 and column 3 data for zone 3}
         	\label{ScenariosAllzone}
         	 \vspace{-0.4em}
\end{figure}

A set of 20 space-time trajectories for three randomly selected days (the same days as those in Fig. \ref{PIAllzones}) for all three zones are depicted in Fig. \ref{ScenariosAllzone}. Trajectories are drawn from the joint Normal distribution as explained in subsection \ref{Trajectory Generation}. As can be seen in the figure, generated scenarios are both temporally and spatially correlated. With temporal correlations, generated scenarios would not evolve through lead-times as a random walk whose magnitudes are formed and limited by marginal densities. Likewise, correlations in space ensures that by looking at a zone the level of uncertainties in the other zones can be estimated.

\begin{figure*}[h]
     \vspace{-1.2em}
     		\begin{center}
     		\centering
     		\noindent
     		\includegraphics[width=10cm]{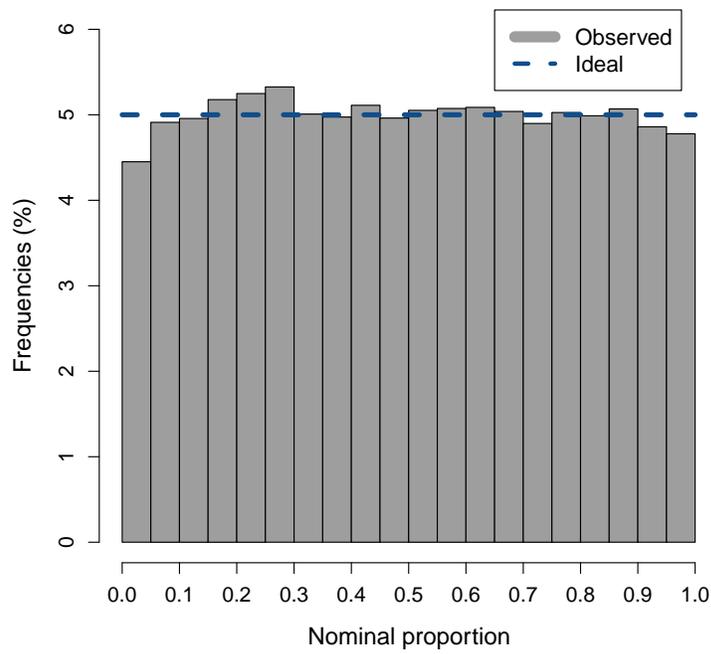}
     	\end{center}
     	\vspace{-1.2em}
     	\caption{ Time trajectories:  evaluation of the probabilistic correctness of generated scenarios with a reliability diagram.  This  diagram gathers the
     	results over whole evaluation dataset  for all look-ahead times for zone 1}
     	\label{PITTime}
     	 \vspace{-0.4em}
 \end{figure*}

\begin{figure}[h]
         \vspace{-1.2em}
 \begin{center}
 \centering
 \noindent
 \includegraphics[width=10cm]{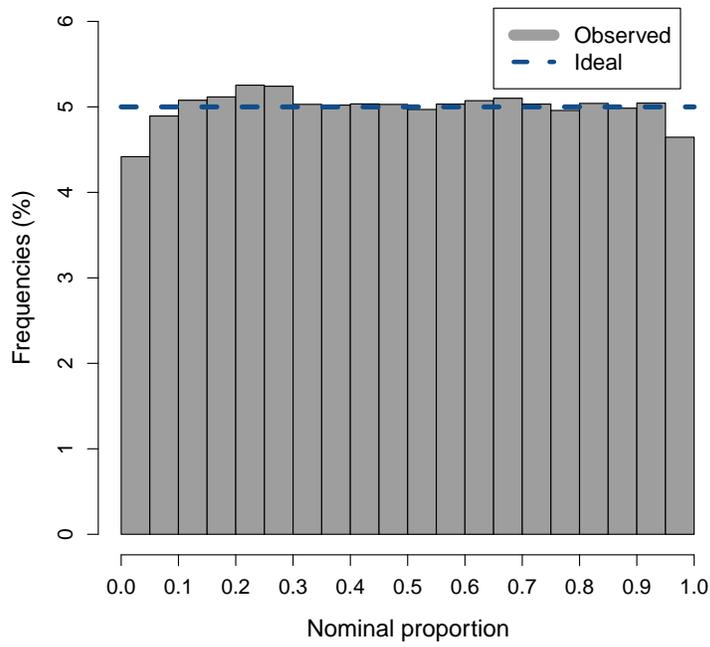}
         	\end{center}
         	\vspace{-1.2em}
         	\caption{ Space-time trajectories: evaluation of the probabilistic correctness of generated scenarios with a PIT histogram. The ideal
         	case of perfect probabilistic correctness is represented by the dash line. This PIT histogram gathers the
         	results for all look-ahead times and zones}
         	\label{PITSpace}
         	 \vspace{-0.4em}
\end{figure}

 Figs.   \ref{PITTime} and \ref{PITSpace} illustrate the reliability diagram and PIT histogram over all evaluation dataset containing all look-ahead times  for temporal  and space-time trajectories respectively. It should be emphasized that these two evaluation criteria measure reliability of trajectories with respect to the marginal distributions. Therefore, in case the later ones as the inputs of the dependency study are not reliable, unreliability would be expanded to the generated trajectories.
  In Figs.  \ref{PITTime} and \ref{PITSpace}, the perfect situation is the case where each bin contains 5 percent of the generated trajectories. As can be seen deviations from the ideal line are very low which  indicate that trajectories are distributed identical to the predictive distributions.    
 
\begin{figure}[h]
   \vspace{-0.6em}
    		\begin{center}
    		\centering
    		\noindent
    		\noindent\makebox[\textwidth]{
    		\includegraphics[width=18cm]{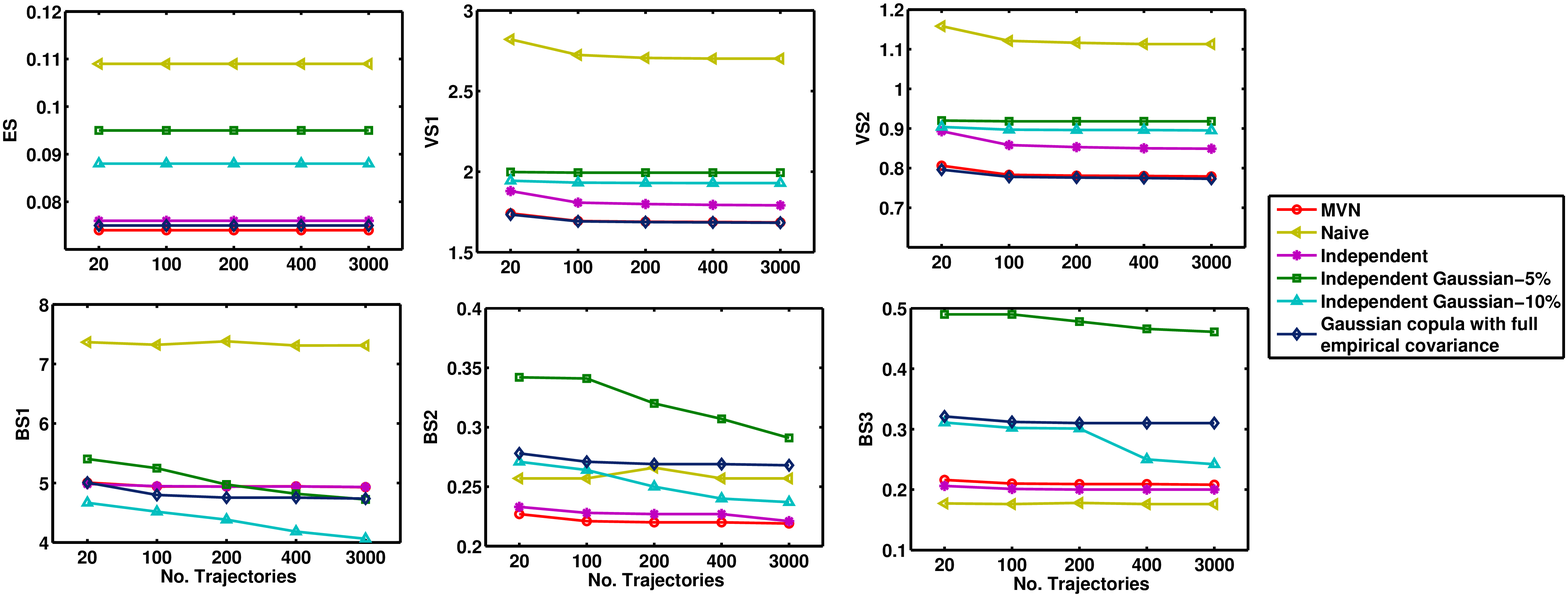}
    		}
    	\end{center}
    	\vspace{-1.2em}
    	\caption{Temporal trajectories: The scores for multivariate normal distribution with recursive covariance and  five other benchmarks. 5 different sizes for the trajectories set are considered. The parameter setting for VS1 is $ w_{ij}=1 \: \forall i,j $, $ \gamma=0.5 $,  for VS2 $ w=$ empirical covariance matrix, $ \gamma=0.5 $. BS1, BS2 and BS3 represent  Brier score for event 1 to 3, respectively. For BS1 and BS3 $ k=11 $ and $ h=4 $ and for BS2   $ k=11 $ and $ h=2 $.}
    	\label{ScoresTime}
    	 \vspace{-0.2em}
\end{figure}

\begin{figure}[h]
    \vspace{-0.9em}
    		\begin{center}
    		\centering
    		\noindent
    		\noindent\makebox[\textwidth]{
    		\includegraphics[width=18cm]{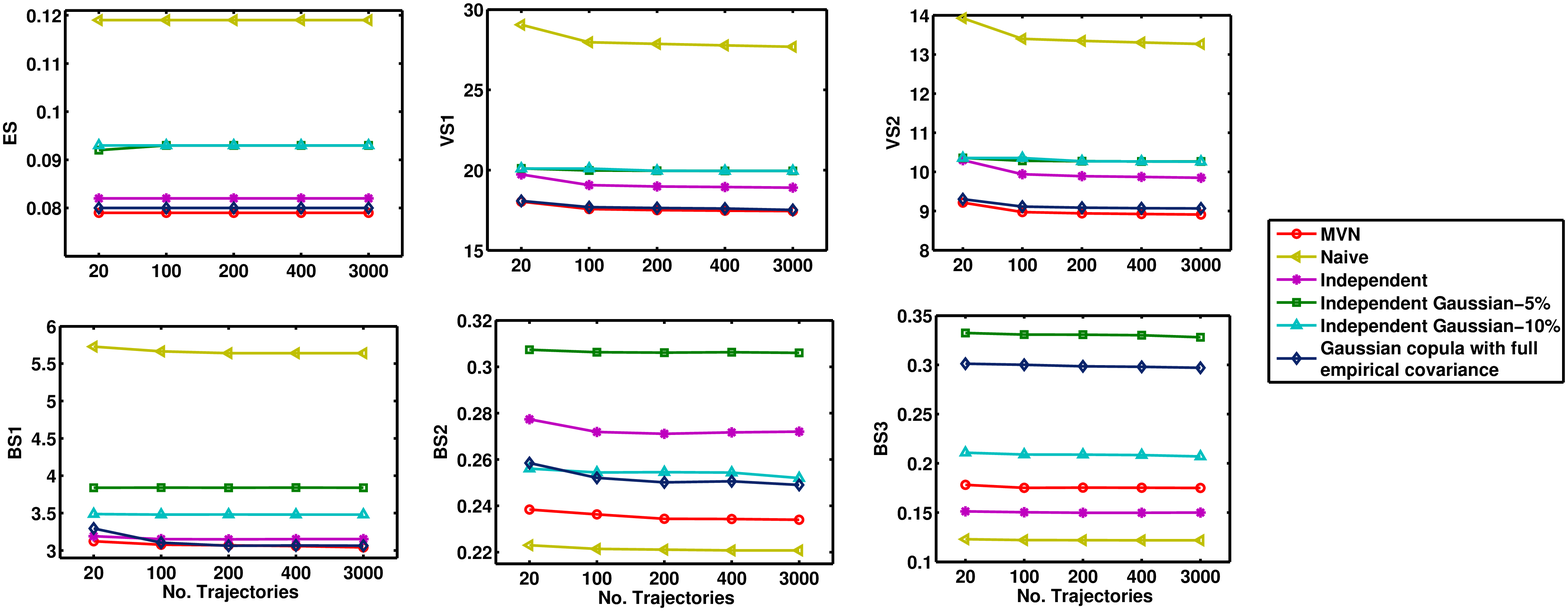}
    		}
    	\end{center}
    	\vspace{-1.2em}
    	\caption{Space-time trajectories: the scores for multivariate normal distribution with recursive covariance and  five other benchmarks. 5 different sizes for the trajectories set are considered. The parameter setting for VS1 is $ w_{ij}=1 \: \forall i,j $, $ \gamma=0.5 $,  for VS2 $ w=$ empirical covariance matrix, $ \gamma=0.5 $. BS1, BS2 and BS3 represent  Brier score for event 1 to 3, respectively. For BS1 and BS3 $ k=11 $ and $ h=4 $ and for BS2   $ k=11 $ and $ h=2 $.}
    	\label{ScoresSpace}
\end{figure}

 The skill of all the methods described in this paper are verified  based on the skill scores explained in section \ref{Scoring Rules} namely energy score (ES), Variogram-based score (VS) and Brier score (BS). To investigate the impact of the number of generated trajectories on discriminating ability of a method, sets of trajectories in 5 different sizes $ S= $ 20, 100, 200, 400 and 3000 are generated. Hereafter, dependency modelling based on multivariate normal distribution with recursive covariance is denoted by MVN. The scores for MVN and  five other benchmarks for time and space-time dependencies are visualized in Figs. \ref{ScoresTime} and \ref{ScoresSpace} respectively. For Variogram-based score, two different settings are assessed.
 For VS1, the weights for all pairs of the components of $ \hat{\textbf{P}}_t $ are set to 1.  Moreover, in order to degrade the impact of pairs of components with relatively lower correlations, in VS2 $ w_{i,j} $ is set to the historical correlation between  components $ i $ and $ j $. As reported in~\cite{scheuerer2015variogram}, $ \gamma $ equal to 0.5 for most cases provides good discriminating ability. Therefore, for both VS1 and VS2, $ \gamma $ is set to 0.5.
   BS1, BS2 and BS3 represent  Brier score for events 1, 2 and 3 explained in subsection \ref{eventscore} respectively. For BS1 and BS3  $ k=11 $ and $ h=4 $ and for BS2  $ k=11 $ and $ h=2$. $ \xi $ for all three events is equal to 0.2. It is to be noted that in all tests for MVN, independent and Gaussian copula with full empirical covariance approaches, the same marginal distributions are used. The score for each experiment is calculated by averaging over 30 independent runs and the values are rounded to 3 digits.
 
 By looking at Figs. \ref{ScoresTime} and \ref{ScoresSpace}, it can be concluded that the impact of number of trajectories varies from one score to another. Increasing the number of trajectories has almost no distinguishable influence on ES when evaluating PV power trajectories. This feature, on one hand can be viewed as an advantage when only limited trajectories are available such as the case with small size ensemble. On the other hand, it can be interpreted as a weakness of this score at discriminating between a very good representation of an predictive density of PV power and a very sparse one. By increasing the number of trajectories, though, VS scores improve continuously for all the four settings. A higher number of trajectories   as a better representative for predictive density can also get a better Brier score, although the improvements are not as noticeable as the case with VS.
 
 
 The skill of independent benchmark  has its lowest difference with that of MVN when the evaluation metric is ES. This supports the conclusion in~\cite{pinson2012evaluating} that ES is weak at discriminating multivariate forecasts when correlations between their components make the only distinction between them.
 
Analysing Figs. \ref{ScoresTime} and \ref{ScoresSpace} reveals that for both time and space-time  dependency, MVN has the best or close to best discriminating ability
to detect highly intermittent time of day, long lasting and gradient intermittency. Gaussian independent method with standard deviation 10\% of the point forecasts presents better skills in comparison with the one with 5\% standard deviation.

 Overall, independent method shows better performance than independent Gaussian. This informs about the benefits of deploying probability forecasting to model predictive distributions of PV generations for each time in the future because normality  assumption is too naive to model variations of PV power.  
 
 The skill of MVN and Gaussian copula with full empirical covariance are quite comparative which means that recursive approach is able to mimic the dependence structure of the multivariate quantity. Therefore, in case the historical data is not available to calculate the empirical covariance matrix, the recursive approach can be a proper alternative.
 
\begin{figure}[h]
    \vspace{-0.9em}
    		\begin{center}
    		\centering
    		\noindent
    		\noindent\makebox[\textwidth]{
    		\includegraphics[width=10cm]{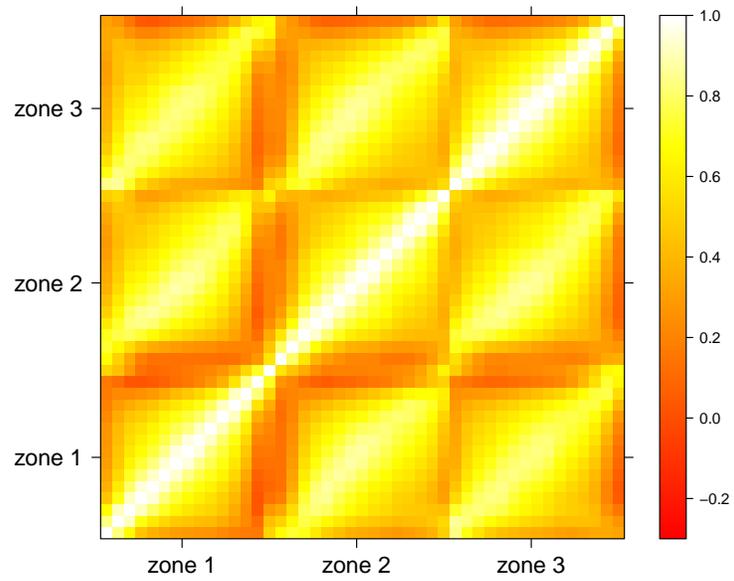}
    		}
    	\end{center}
    	\vspace{-1.2em}
    	\caption{Visualization of correlation matrix for the transformed measurements on the time grid, the matrix is of size of 45$ \times $ 45. Red to white colours represent the lowest to the highest correlations}
    	\label{corrSpace}
\end{figure}

 Analysing the skill of independent benchmark and MVN can reveal the importance of modelling dependency structures. While independent benchmark takes advantage of probabilistic forecasts, it treats each look-ahead time $ k $ and location $ z $ independently. However,  intermittent generations are dependent both in time and space. To get an idea about the significance of dependency of PV generations in time and space, the empirical covariance matrix for the transformed observations on the time grid is visualized in Fig. \ref{corrSpace}. According to this figure, time dependency decreases as the difference between lead times increases. Furthermore, there is high dependency in space specially for adjacent lead-times. As the generated trajectories are to be used in operational problems, providing miscalibrated forecasts based on wrong assumption of independency leads to suboptimal decisions.
 
 Although independent benchmark is insufficiently skillful with respect to MVN, it provides much better ES in comparison with Gaussian independent methods. In terms of VS, again independent approach surpasses Gaussian independent  benchmarks.
 This can validate that probabilistic forecasts can offer more accurate information about the uncertainties involved than the naive assumption of normality distribution for forecast errors.  

\begin{figure}[h]
   \vspace{-0.6em}
    		\begin{center}
    		\centering
    		\noindent
    		\noindent\makebox[\textwidth]{
    		\includegraphics[width=15cm]{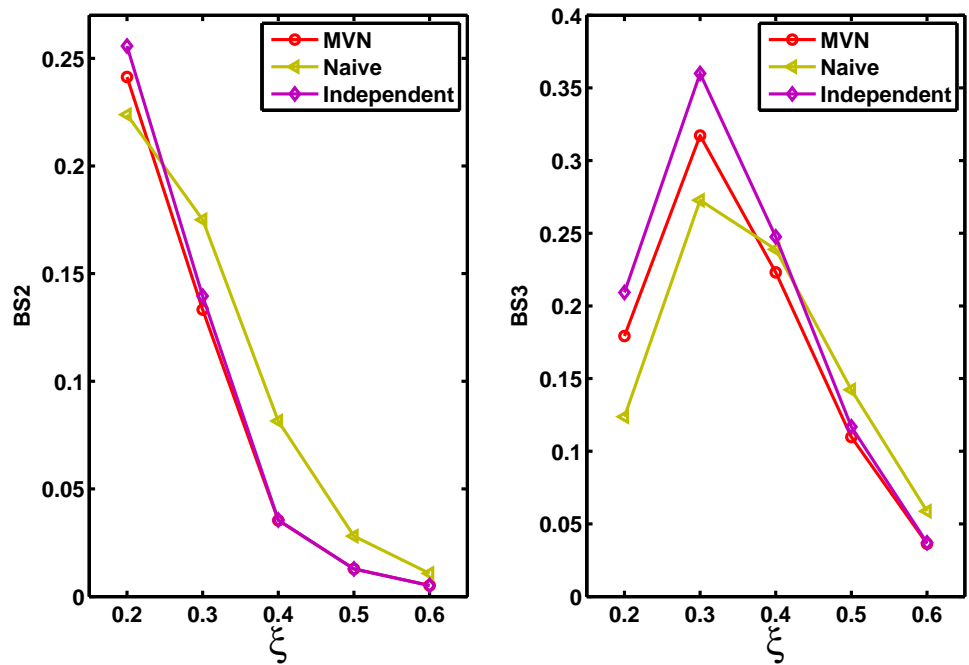}
    		}
    	\end{center}
    	\vspace{-1.2em}
    	\caption{Space-time trajectories: BS2 and BS3 scores for MVN, Naive and Independent methods for five different values of $ \xi$.   For  BS3 $ k=11 $ and $ h=4 $ and for BS2   $ k=11 $ and $ h=2 $.}
    	\label{NaiveXi}
    	 \vspace{-0.2em}
\end{figure}

According to Fig.  \ref{ScoresSpace}, the naive method shows the best BS2 and BS3 scores. However, further investigations showed that as the $ \xi $ increases (toward the rare events), MVN gets better BS2 and BS3 in comparison with the naive method. Fig. \ref{NaiveXi} illustrates BS2 and BS3  for the MVN, naive and independent methods for five different values of $ \xi$. As can be seen for $ \xi $ more than $ 0.2 $,  BS2 scores given by the MVN and independent methods are better than those given by the naive method. The same results obtained for BS3 when $ \xi $ is more than $ 0.3 $. It can be concluded that the scenarios generated by the naive method are able to describe the frequent events however, they are weak at informing about the less frequent or rare events.
 
\section{Conclusions}
As there is always a level of uncertainty attached to the PV generation forecasting, uncertainty quantification is an integral factor to maintain acceptable levels of reliability, security and profitability in power systems. Due to inertia of meteorological systems,  prediction errors are correlated in time and space. In this study, joint predictive distributions based on marginal densities are modelled and space-time trajectories describing evolution of forecast errors through successive lead-times and locations are generated. Several scoring rules are used to measure the similarity of different set of forecasts with respect to the PV generation observations. Three events are proposed which cater for the specific case of PV generations. These events measure the level of intermittency in PV generations with respect to the maximum expected PV power for each time of day. Analysing the results validates the superiority of the probabilistic forecasts in describing the uncertainties over the common practice of normality assumption for forecast errors. Moreover,  trajectories drawn from multivariate distributions present better skill in almost all the experiments. Therefore,  in operational problems which are sensitive to the dependency structures, one may  conclude that using probabilistic forecasts when space-time correlations are neglected may provide only a suboptimal solution. Our investigation verified that in case that historical data is not available to find dependency structures of PV generations, modelling covariance matrix recursively can be viewed as an alternative approach. While in this paper, multivariate distribution  is assumed to be Gaussian, further investigations are required to find the best copula function which can describe the behaviour of the PV generations in time and space. Therefore, in our future work, we will try to perform various  tests to evaluate Archimedean, empirical, Frank, T, Clayton, Gumbel copulas and find the right one for the case of PV generations.
\section{References}
  \bibliographystyle{elsarticle-num} 
  \bibliography{reftest}

\begin{thebibliography}{10}
\expandafter\ifx\csname url\endcsname\relax
  \def\url#1{\texttt{#1}}\fi
\expandafter\ifx\csname urlprefix\endcsname\relax\def\urlprefix{URL }\fi
\expandafter\ifx\csname href\endcsname\relax
  \def\href#1#2{#2} \def\path#1{#1}\fi

\bibitem{inman2013solar}
R.~H. Inman, H.~T. Pedro, C.~F. Coimbra, Solar forecasting methods for
  renewable energy integration, Progress in Energy and Combustion Science
  39~(6) (2013) 535--576.

\bibitem{marquez2013forecasting}
R.~Marquez, V.~G. Gueorguiev, C.~F. Coimbra, Forecasting of global horizontal
  irradiance using sky cover indices, Journal of Solar Energy Engineering
  135~(1) (2013) 011017.

\bibitem{chu2013hybrid}
Y.~Chu, H.~T. Pedro, C.~F. Coimbra, Hybrid intra-hour dni forecasts with sky
  image processing enhanced by stochastic learning, Solar Energy 98 (2013)
  592--603.

\bibitem{mellit2006adaptive}
A.~Mellit, M.~Benghanem, S.~Kalogirou, An adaptive wavelet-network model for
  forecasting daily total solar-radiation, Applied Energy 83~(7) (2006)
  705--722.

\bibitem{quan2015computational}
H.~Quan, D.~Srinivasan, A.~M. Khambadkone, A.~Khosravi, A computational
  framework for uncertainty integration in stochastic unit commitment with
  intermittent renewable energy sources, Applied Energy 152 (2015) 71--82.

\bibitem{lorenz2009irradiance}
E.~Lorenz, J.~Hurka, D.~Heinemann, H.~G. Beyer, Irradiance forecasting for the
  power prediction of grid-connected photovoltaic systems, Selected Topics in
  Applied Earth Observations and Remote Sensing, IEEE Journal of 2~(1) (2009)
  2--10.

\bibitem{mathiesen2013geostrophic}
P.~Mathiesen, J.~M. Brown, J.~Kleissl, Geostrophic wind dependent probabilistic
  irradiance forecasts for coastal california, Sustainable Energy, IEEE
  Transactions on 4~(2) (2013) 510--518.

\bibitem{chu2015real}
Y.~Chu, M.~Li, H.~T. Pedro, C.~F. Coimbra, Real-time prediction intervals for
  intra-hour dni forecasts, Renewable Energy 83 (2015) 234--244.

\bibitem{alessandrini2015analog}
S.~Alessandrini, L.~Delle~Monache, S.~Sperati, G.~Cervone, An analog ensemble
  for short-term probabilistic solar power forecast, Applied Energy 157 (2015)
  95--110.

\bibitem{papaefthymiou2006integration}
G.~Papaefthymiou, P.~Schavemaker, L.~Van~der Sluis, W.~Kling, D.~Kurowicka,
  R.~Cooke, Integration of stochastic generation in power systems,
  International Journal of Electrical Power \& Energy Systems 28~(9) (2006)
  655--667.

\bibitem{yang2013solar}
D.~Yang, C.~Gu, Z.~Dong, P.~Jirutitijaroen, N.~Chen, W.~M. Walsh, Solar
  irradiance forecasting using spatial-temporal covariance structures and
  time-forward kriging, Renewable Energy 60 (2013) 235--245.

\bibitem{qin2013incorporating}
Z.~Qin, W.~Li, X.~Xiong, Incorporating multiple correlations among wind speeds,
  photovoltaic powers and bus loads in composite system reliability evaluation,
  Applied Energy 110 (2013) 285--294.

\bibitem{gueymard2011assessment}
C.~A. Gueymard, S.~M. Wilcox, Assessment of spatial and temporal variability in
  the us solar resource from radiometric measurements and predictions from
  models using ground-based or satellite data, Solar Energy 85~(5) (2011)
  1068--1084.

\bibitem{yang2012novel}
C.~Yang, L.~Xie, A novel arx-based multi-scale spatio-temporal solar power
  forecast model, in: 2012 North American Power Symposium (NAPS), 2012, pp.
  1--6.

\bibitem{widen2011correlations}
J.~Wid{\'e}n, Correlations between large-scale solar and wind power in a future
  scenario for sweden, Sustainable Energy, IEEE Transactions on 2~(2) (2011)
  177--184.

\bibitem{yang2015multitime}
C.~Yang, A.~Thatte, L.~Xie, et~al., Multitime-scale data-driven spatio-temporal
  forecast of photovoltaic generation, Sustainable Energy, IEEE Transactions on
  6~(1) (2015) 104--112.

\bibitem{bessa2015spatial}
R.~J. Bessa, A.~Trindade, V.~Miranda, Spatial-temporal solar power forecasting
  for smart grids, Industrial Informatics, IEEE Transactions on 11~(1) (2015)
  232--241.

\bibitem{berdugo2011analog}
V.~Berdugo, C.~Chaussin, L.~Dubus, G.~Hebrail, V.~Leboucher, Analog method for
  collaborative very-short-term forecasting of power generation from
  photovoltaic systems, Next Generation Data Mining Summit: Ubiquitous
  Knowledge Discovery for Energy Management in Smart Grids and Intelligent
  Machine-to-Machine (M2M) Telematics.

\bibitem{papaefthymiou2009using}
G.~Papaefthymiou, D.~Kurowicka, Using copulas for modeling stochastic
  dependence in power system uncertainty analysis, Power Systems, IEEE
  Transactions on 24~(1) (2009) 40--49.

\bibitem{pinson2009probabilistic}
P.~Pinson, H.~Madsen, H.~A. Nielsen, G.~Papaefthymiou, B.~Kl{\"o}ckl, From
  probabilistic forecasts to statistical scenarios of short-term wind power
  production, Wind energy 12~(1) (2009) 51--62.

\bibitem{gneiting2007probabilistic}
T.~Gneiting, F.~Balabdaoui, A.~E. Raftery, Probabilistic forecasts, calibration
  and sharpness, Journal of the Royal Statistical Society: Series B
  (Statistical Methodology) 69~(2) (2007) 243--268.

\bibitem{pinson2012evaluating}
P.~Pinson, R.~Girard, Evaluating the quality of scenarios of short-term wind
  power generation, Applied Energy 96 (2012) 12--20.

\bibitem{scheuerer2015variogram}
M.~Scheuerer, T.~M. Hamill, Variogram-based proper scoring rules for
  probabilistic forecasts of multivariate quantities, Monthly Weather Review
  143~(4) (2015) 1321--1334.

\bibitem{koenker1978regression}
R.~Koenker, G.~Bassett~Jr, Regression quantiles, Econometrica: journal of the
  Econometric Society (1978) 33--50.

\bibitem{website55}
\href{https://crowdanalytix.com}{Global energy forecasting competition 2014
  probabilistic solar power forecasting}.
\newline\urlprefix\url{https://crowdanalytix.com}

\bibitem{gneiting2007strictly}
T.~Gneiting, A.~E. Raftery, Strictly proper scoring rules, prediction, and
  estimation, Journal of the American Statistical Association 102~(477) (2007)
  359--378.

\bibitem{sklar1959fonctions}
M.~Sklar, Fonctions de r{\'e}partition {\`a} n dimensions et leurs marges,
  Universit{\'e} Paris 8, 1959.

\bibitem{nelsen2013introduction}
R.~B. Nelsen, An introduction to copulas, Vol. 139, Springer Science \&
  Business Media, 2013.

\bibitem{demarta2005t}
S.~Demarta, A.~J. McNeil, The t copula and related copulas, International
  statistical review 73~(1) (2005) 111--129.

\bibitem{louie2014evaluation}
H.~Louie, Evaluation of bivariate archimedean and elliptical copulas to model
  wind power dependency structures, Wind Energy 17~(2) (2014) 225--240.

\bibitem{diaz2014simulation}
G.~D{\'\i}az, P.~G. Casielles, J.~Coto, Simulation of spatially correlated wind
  power in small geographic areas—sampling methods and evaluation,
  International Journal of Electrical Power \& Energy Systems 63 (2014)
  513--522.

\bibitem{diaz2014note}
G.~D{\'\i}az, A note on the multivariate archimedean dependence structure in
  small wind generation sites, Wind Energy 17~(8) (2014) 1287--1295.

\bibitem{pinson2007non}
P.~Pinson, H.~A. Nielsen, J.~K. M{\o}ller, H.~Madsen, G.~N. Kariniotakis,
  Non-parametric probabilistic forecasts of wind power: required properties and
  evaluation, Wind Energy 10~(6) (2007) 497--516.

\bibitem{bacher2009online}
P.~Bacher, H.~Madsen, H.~A. Nielsen, Online short-term solar power forecasting,
  Solar Energy 83~(10) (2009) 1772--1783.

\bibitem{Batch2015}
F.~Golestaneh, H.~Gooi, Batch and sequential forecast models for photovoltaic
  generationl, in: IEEE PES 2015 General Meeting, Denver, Colorado, USA, 2015.

\end{thebibliography}

     
 \begin{figure}[!t]
  	\vspace{-0.4em}
  	\centering
  	\includegraphics[width=8cm, height=7cm]{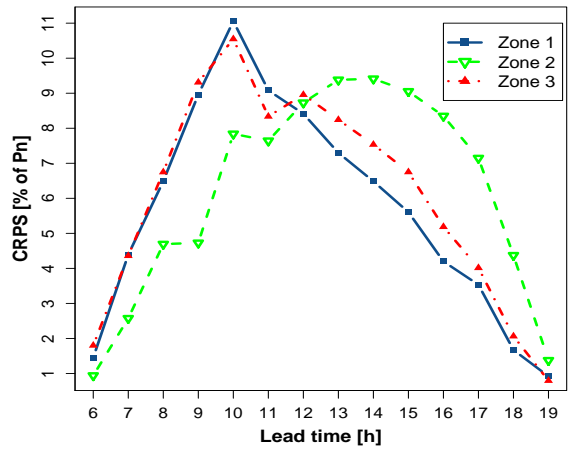}
  	\caption{Skill evaluation of the probabilistic forecasts of PV generation for three contiguous PV zones as a function of forecast horizon}
  	\label{CRPSFi}
  	\vspace{-0.6em}
  \end{figure}


%
%
%
\end{document}